\documentclass[aps,prl,reprint,superscriptaddress]{revtex4-1}

\usepackage{graphicx}
\usepackage{fixltx2e}

\begin{document}


\title{\Large Imaging the breakdown of molecular-frame dynamics through rotational uncoupling}


\author{Lucas J. Zipp}
\affiliation{Stanford PULSE Institute, SLAC National Accelerator Laboratory, Menlo Park, California 94025, USA}
\affiliation{Department of Physics, Stanford University, Stanford, California 94305, USA}

\author{Adi Natan}
\affiliation{Stanford PULSE Institute, SLAC National Accelerator Laboratory, Menlo Park, California 94025, USA}

\author{Philip H. Bucksbaum}

\affiliation{Stanford PULSE Institute, SLAC National Accelerator Laboratory, Menlo Park, California 94025, USA}
\affiliation{Department of Physics, Stanford University, Stanford, California 94305, USA}
\affiliation{Department of Applied Physics, Stanford University, Stanford, California 94305, USA}



\begin{abstract}
We demonstrate the breakdown of molecular-frame dynamics induced by the uncoupling of molecular rotation from electronic motion in molecular Rydberg states. We observe this non-Born-Oppenheimer regime in the time domain through photoelectron imaging of a coherent molecular Rydberg wave packet in N\textsubscript{2}. The photoelectron angular distribution shows a radically different time evolution than that of a typical molecular-frame-fixed electron orbital, revealing the uncoupled motion of the electron as it precesses around the \textit{averaged} anisotropic potential of the rotating ion-core.
\end{abstract}


\maketitle


 In the standard Born-Oppenheimer picture, electrons occupy orbitals that are fixed to the molecular frame and display the symmetries of the underlying molecular structure. Recent experiments in ultrafast and strong field physics have utilized aligned ensembles of molecules, allowing for molecular frame measurements of excited state dynamics \cite{bisgaard_2009,hockett_2011,qin_direct_2011,gesner_femtosecond_2006} and XUV and tunnel ionization \cite{kelkensberg_xuv_2011,williams_imaging_2012,litvinyuk_2003,meckel_signatures_2014,staudte_angular_2009,holmegaard_photoelectron_2010} that are sensitive to electron orbital geometry. This simple picture breaks down for non-penetrating Rydberg electrons, where the electron wave function has minimal overlap with the ion-core. Coriolis-type forces can decouple the electron motion from the rotating core leading to a breakdown of Born-Oppenheimer molecular frame dynamics in a process known as $l$-uncoupling \cite{mulliken_rydberg_1964}.

The complex interplay between electronic and nuclear motion in the $l$-uncoupling regime presents a unique opportunity to study non-Born-Oppenheimer rotational-electronic coupling in molecules. The phenomenon of $l$-uncoupling has been previously inferred by the perturbed spacing of rotational levels of high-lying Rydberg states \cite{herzberg_high_1982,chang_observation_2005,huber_f_1994,jungen_2003,jungen_absorption_1969,mccormack_analysis_1991}, but the dynamics have never before been observed directly.

Here we report the direct imaging in the time domain of the uncoupled motion of a molecular Rydberg electron. This is achieved through multiphoton preparation and subsequent photoelectron imaging of a coherent superposition of electronic states in the 4$f$ Rydberg manifold of N\textsubscript{2}. We track the angular motion of the $l$-uncoupled Rydberg electron and measure the effect of uncoupling on its laboratory-frame dynamics, providing a close view of the coherent dynamics of a molecular system in a non-Born-Oppenheimer regime. This work complements previous angle-integrated measurements of Rydberg molecules that have focused on other aspects of rotational-electronic coupling, most notably stroboscopic effects on the radial motion of a Rydberg wave packet \cite{labastie_stroboscopic_1984,smith_role_2003,li_homonuclear_2011,fielding_rydberg_2005}.
     
\begin{figure*}
\includegraphics[trim = 0mm 0mm 0mm 0mm, clip,width=\textwidth]{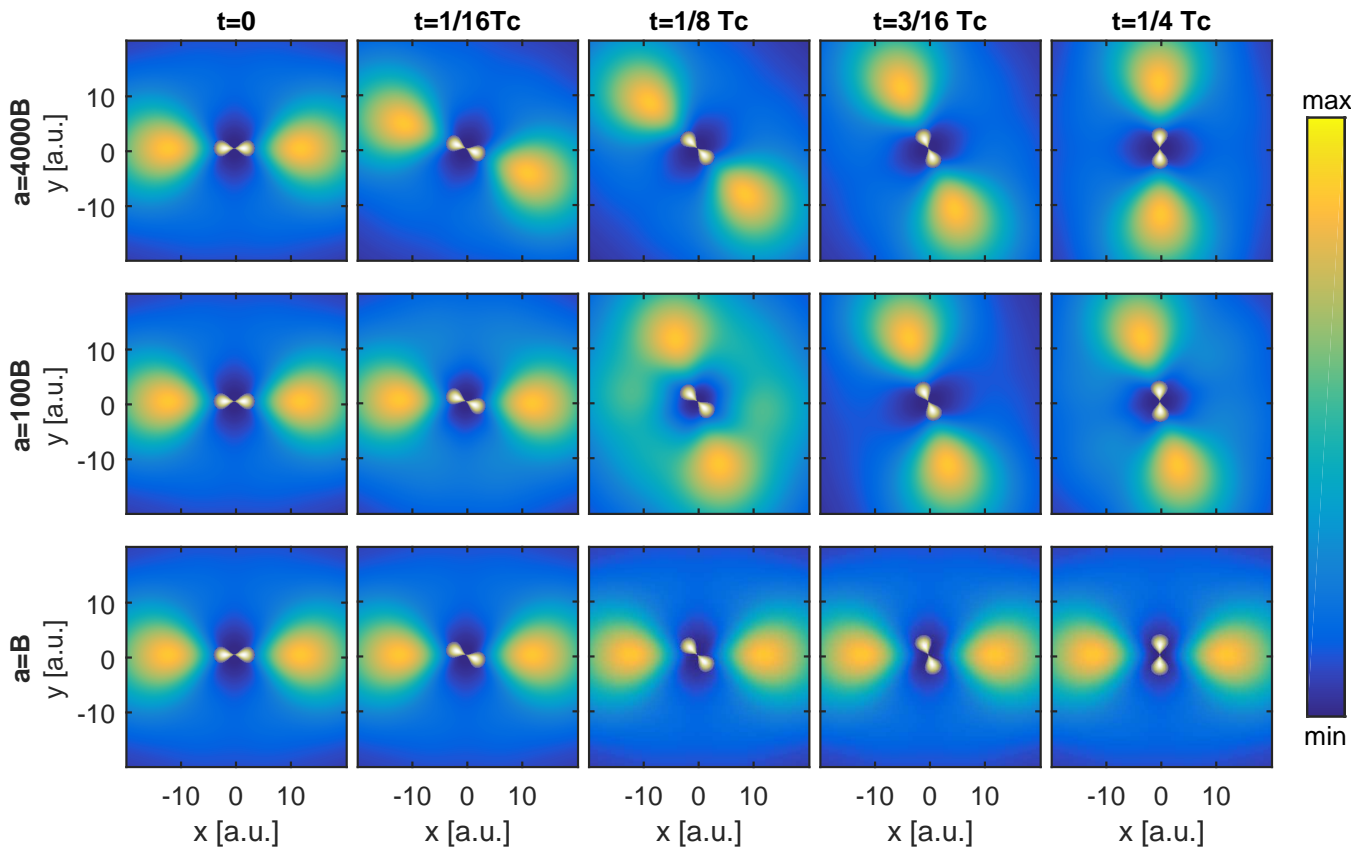}
 \caption{\label{Fig1} Breakdown of molecular frame dynamics. The outer lobes (blue-yellow color scale) correspond to the Rydberg electron density with a 4$f$ orbital radial distribution, while the inner lobe (white) shows the angular distribution of the nuclei, with an artificially chosen radial distribution for better visualization. The simulated homonuclear diatomic system starts out in a localized rotational wave packet centered at $|J=30, M=30\rangle$ with an $f$ Rydberg electron aligned along the internuclear axis in a  $\Lambda = 0$ state. The electron and ion-core position densities are plotted at times corresponding to fractions of the classical nuclear rotation period, $T_c=\pi/(B\sqrt{J(J+1)})$. Each row displays the dynamics for a different value of the electronic  splitting parameter $a$, as given in equation (2). The top row corresponds to the Born-Oppenheimer regime, while the bottom row shows completely uncoupled nuclear and rotational dynamics. (See Supplemental Material for full movies.)}
 \end{figure*}

The $nl$ Rydberg manifold of a molecule consists of $(2l+1)$ states, which in the Born-Oppenheimer limit correspond to the quantized projections $\Lambda$ of the electronic orbital angular momentum onto the internuclear axis. The coupling between the various angular momenta of a diatomic molecule can be characterized with basis sets known as Hund's cases \cite{hougen_1970}. In the case of the singlet Rydberg states of N\textsubscript{2}, with electronic spin $S=0$, the transition to the $l$-uncoupling regime is then described as a change in basis from Hund's case (b) (the Born-Oppenheimer limit) to Hund's case (d) (the uncoupled limit). Excluding nuclear spin degrees of freedom, this corresponds to a change in quantum labels from $|JM;l \Lambda \rangle \rightarrow |JM;lR\rangle$, where $J$ is the total angular momentum of the system with lab frame projection $M$, $l$ is the orbital angular momentum of the Rydberg electron, and $R$ is the total angular momentum of the ion-core. In the fully uncoupled limit, where the eigenstates of the system are given by Hund's case (d) states, the electron wave function is totally decoupled from the ion-core molecular frame. 

The full rotational-electronic Hamiltonian of a diatomic molecule is given by \cite{brown_electronic_2003}:
\begin{equation} \label{full_H}
H=\hat{H}_{ev}+B\hat{R}^2,
 \end{equation}
where $\hat{H}_{ev}$ is the vibronic Born-Oppenheimer Hamiltonian and $\hat{R}$ corresponds to the rotational angular momentum of the nuclei with rotational constant $B$. $\hat{H}_{ev}$ is diagonal in the Born-Oppenheimer Hund's case (b) basis set, and for non-penetrating Rydberg states, this energy is approximately:
\begin{equation} \label{simplified_H}
{\langle \hat{H}_{ev} \rangle}_{nl\Lambda} = E_{nl} +a \Lambda^2,
 \end{equation}
where $E_{nl}$ is the non-rotating energy (electronic and vibrational) of the $|nl\, \Lambda=0 \rangle$ Rydberg state and $a$ is a constant that depends on the strength of the anisotropic interaction of the Rydberg electron with the core \cite{jungen_absorption_1969}. Non-penetrating Rydberg states refers to Rydberg electrons with minimal overlap with the ion-core wave function, as is generally true for Rydberg electrons with $l>2$ \cite{eyler_1986}. When the electronic splitting is large relative to the rotational energy spacing of the system, which occurs when $a \Lambda^2 >> BR^2$, then the Born-Oppenheimer approximation holds, and the electron wave function is firmly fixed to the molecular frame. As the electronic energy splitting between $\Lambda$ states decreases relative to the rotational energy, the electronic motion and the nuclear rotation begin to mutually perturb each other. This results in the well-known behavior of $\Lambda$-doubling, which removes the degeneracy between even and odd parity states for $\Lambda \neq 0$ \cite{mulliken_rydberg_1964}. If the electronic splitting is reduced further, the $l$-uncoupling regime is reached and the electron orbital angular momentum uncouples from the molecular axis and $\Lambda$ is no longer a good quantum number. Instead, the eigenstates are characterized by the projection of the Rydberg electron orbital angular momentum onto the \textit{rotational} axis of the core. The direct consequence is that by exciting a coherent superposition of sublevels in an $l$-uncoupled Rydberg manifold, the electron wave packet oscillates in the \textit{time-average} molecular potential created by the rotating molecule. The frequency of these angular oscillations depends on the strength of the interaction of the Rydberg electron with the anisotropic ion-core potential and is independent of the rotational period of the core.

 \begin{figure*}
 \includegraphics[trim = 0mm 0mm 0mm 0mm, clip, width=\textwidth]{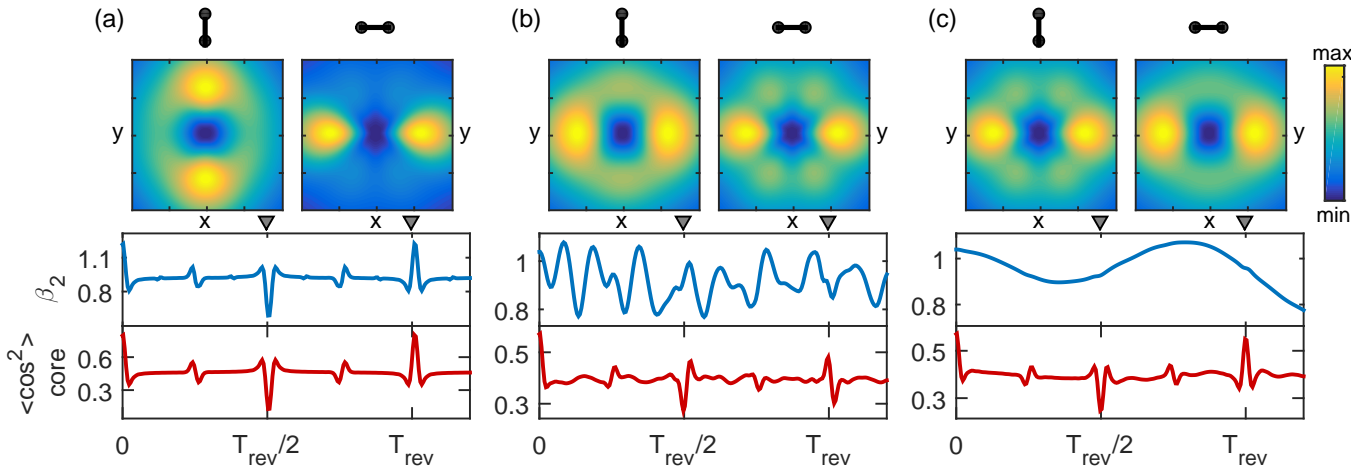}
 \caption{\label{Fig2}The simulated electron and ion-core angular distribution time evolution of a $nf$ Rydberg electron. A model N\textsubscript{2}-like diatomic molecule is pumped to an $nf$ Rydberg state with a 5-photon excitation and subsequently probed through single photon ionization (linear polarization of pump and probe along the x-axis). The upper row of images show the electron position density of the Rydberg electron (before probing) at times corresponding to peak ion-core alignment and anti-alignment which occur near T\textsubscript{rev} and T\textsubscript{rev}/2. A 4$f$ hydrogenic radial wave function is used for visualization. The molecular alignment direction for each image is depicted by the ball-and-stick graphic. The $\beta_2$ paramater of the photoelectron angular distribution and the $<cos^2>$ alignment value of the ion-core distribution is also shown. (a) Simulation for multiphoton excitation restricted to the $\Lambda=0$ eigenstate of the $f$ manifold with $a=2400B$, where $B$ is the rotational constant of the cation core. The PAD displays a typical Born-Oppenheimer molecular alignment signal. (b-c) Simulation for multiphoton excitation assuming all levels of the f manifold are accessible, with an electronic splitting of  $a=6B$ and $a=0.6B$, showing the progressive development of $l$-uncoupled dynamics.}
 \end{figure*}
 
The transition from the Born-Oppenheimer regime to the $l$-uncoupled regime is accompanied by dramatic changes in the coupled electronic-nuclear angular dynamics. This can be visualized in the case of quasi-classical nuclear rotational wave packets, similar to those achievable with the optical centrifuge technique \cite{korobenko_adiabatic_2016}. Using the Hamiltonian in Eq. (\ref{simplified_H}), we simulate the electron and ion-core position density as a function of time (see Supplementary Material for details). The Rydberg electron is initially aligned along the internuclear axis of the rotating molecule, and the ensuing dynamics of the system for several values of the electronic splitting parameter $a$ are shown in Fig. \ref{Fig1}. For large electronic splittings, the system follows Born-Oppenheimer dynamics, and the electronic wave function is tightly bound to the rotating internuclear axis (top row). As the interaction strength decreases between the Rydberg electron and the anisotropic part of the core potential, the electron motion first lags behind the core rotation before jumping ahead, producing an oscillatory motion resembling loosely coupled pendula (middle row). Further decrease in the anisotropic interaction strength leads to uncoupled motion of the electron and the core (bottom row). The nuclei continue to rotate, while the Rydberg electron density remains fixed in the lab reference frame, unperturbed by the nuclear motion.

The uncoupling behavior of Rydberg states can be observed directly in time- and angle-resolved photoelectron spectra. A multiphoton pump pulse initiates the $l$-uncoupling dynamics by exciting a molecular Rydberg manifold coupled to a coherent rotational wave packet. The state of the system may then be probed at later times through single photon ionization of the Rydberg state. Through numerical simulations we can explore how the time-dependent photoelectron angular distribution (PAD) in this pump-probe scheme is expected to change with the transition from Born-Oppenheimer dynamics to the $l$-uncoupled regime. The simulations use the same Hamiltonian of Eq. (\ref{simplified_H}) used to generate the visualization in Fig. \ref{Fig1}, however we now prepare the Rydberg wave packet assuming an impulsive, perturbative five-photon transition from the ground state to a 4$f$ Rydberg manifold, and include an impulsive single photon ionization step to create the photoelectron angular distribution. The PAD is then characterized by a sum of even-order Legendre polynomials,
\begin{equation}
 I(\theta)=\frac{\sigma}{4 \pi}[1+\beta _2 P_2 (\cos\theta) + \beta _4 P_4 (\cos\theta) + \cdots], 
 \end{equation}
where $\sigma$ is the total angle-integrated yield.

Figure \ref{Fig2} shows the simulated PAD for several values of the splitting parameter $a$. For excitation to an isolated substate of a manifold with large electronic splittings, the time-dependent PAD exhibits a typical Born-Oppenheimer molecular frame alignment dependence (Fig. \ref{Fig2}(a)). As the electronic splitting decreases and all the manifold states are coherently populated (Fig. \ref{Fig2}(b)), the $l$-uncoupling regime is approached and the PAD alignment signal no longer follows the \textit{instantaneous} ion-core alignment, and rotational-electronic coupling significantly perturbs the dynamics of the ion-core rotational wave packet. For very small electronic splittings (Fig. \ref{Fig2}(c)), a clear separation of the rotational and electronic time scales is achieved in what is sometimes referred to as a reverse Born-Oppenheimer regime.
 
 We directly observe this dynamic $l$-uncoupling behavior in the 4$f$ Rydberg manifold of N\textsubscript{2}. We employ a $\approx$100 fs 400 nm excitation pulse with a peak intensity of $\approx$10\textsuperscript{14} W/cm\textsuperscript{2}. The pulse excites molecules to the 4$f$ Rydberg manifold via a transient five-photon resonance due to the intensity-dependent ac Stark shift \cite{jr_stark-assisted_2007}. The electron wave packet is subsequently probed through single photon ionization by a time-delayed, co-propagating 800 nm pulse of similar duration and intensity as the pump pulse. The resulting photoelectron momentum distribution is collected using a velocity map imaging spectrometer (VMI)\cite{eppink_1997}. The pulses have linear polarizations parallel to each other and to the face of the microchannel plate. More information on the experimental setup can be found elsewhere \cite{zipp_probing_2014}.

The photoelectron angular distribution of the 4$f$ Rydberg state is extracted from the momentum distribution by first subtracting the background pump-only ionization signal and then integrating over the energy width of the photoelectron peak.
 \begin{figure}
 \includegraphics[trim = 2mm 14mm 46mm 0mm, clip,width=\columnwidth]{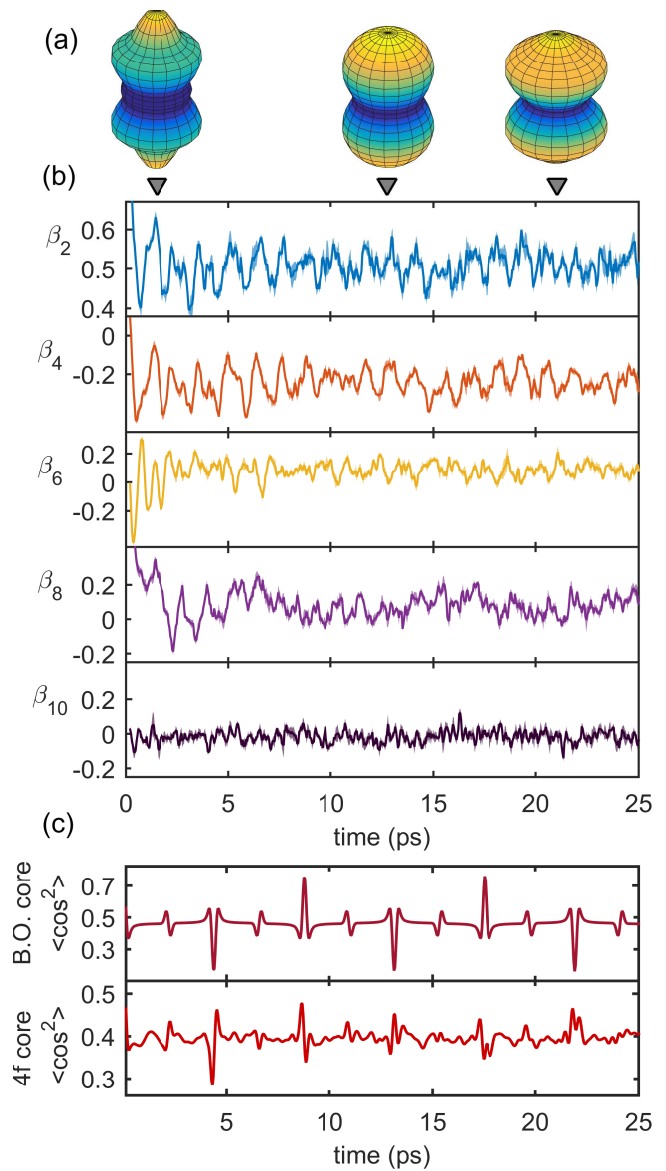}
 \caption{\label{Fig3}Direct observation of $l$-uncoupling dynamics. (a) Polar plots of the experimentally extracted photoelectron angular distribution from the 4$f$ Rydberg manifold in N\textsubscript{2} at selected probe times. (b) $\beta$ values extracted from a least-squares fit of the data to equation (3). The error is plotted as a shaded region (where larger than the line thickness) corresponding to the standard error from three separate pump-probe scans. (c) Simulated ion-core alignment after multiphoton excitation in N\textsubscript{2}. The top plot corresponds to excitation to a single $\Sigma$ state in the Born-Oppenheimer regime, while the lower plot shows the expected ion-core alignment signal for excitation to the 4$f$ Rydberg manifold.}
 \end{figure}
 
The experimentally measured $\beta _n$ values of the photoelectron distribution for room temperature N\textsubscript{2} as a function of pump-probe delay are shown in Fig. \ref{Fig3}(b). The magnitude of $\beta _n$ for $n>8$ is negligible, as can be seen for $\beta_{10}$. This agrees with the identification of the Rydberg state as an $f$ orbital since single-photon ionization selection rules dictate a maximum partial wave of $l=4$, and hence a maximum nonzero $\beta$ order of $\beta _8$ for an $f$ orbital. Clear oscillatory signatures are seen in the $\beta$ parameters, some of which survive longer than the full 30 ps scan range. The form and magnitude of these angular oscillations are shown in the polar plots in Fig. 3(a).

Although the  initial multiphoton excitation must also coherently populate nuclear rotational states in N\textsubscript{2}\textsuperscript{+}, the time-dependent PAD shows no direct correlation with the expected ion-core alignment dynamics of N\textsubscript{2}\textsuperscript{+}. The simulated ion-core alignment for the multiphoton excited 4$f$ state is shown in Fig. 3(c) along with the typical Born-Oppenheimer signal for comparison. The measured time-dependent PAD shows no evidence of the half and full rotational revival periods near 4.3 and 8.7 ps, which are the expected signatures of coherent rotational wave packets in Born-Oppenheimer molecular systems \cite{qin_direct_2011,tsubouchi_photoelectron_2001}. Instead, the quantum beats in the angular distribution correspond to the small electronic splittings of the 4$f$ manifold in the presence of the anisotropic core. In the completely uncoupled regime described by Hund's case (d),  selection rules dictate that the state of the core is unchanged during photoionization and hence different ion-core rotational states are not coupled together in the photoionization process \cite{xie_selection_1990}. The result is a PAD which is entirely insensitive to the molecular frame motion. This is in contrast to the excitation of a low-lying valence electronic state, where the electron wave function follows the instantaneous orientation of the molecular frame and the time-dependent PAD exhibits large variation in the $\beta$ parameters only near the quantum revivals of the rotational wave packet \cite{qin_direct_2011,tsubouchi_photoelectron_2001}.

 \begin{figure}
 \includegraphics[trim = 0mm 0mm 0mm 0mm, clip,width=\columnwidth]{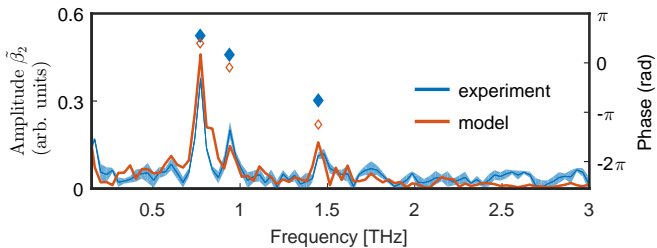}
 \caption{\label{Fig4} Fourier analysis. The amplitude of the discrete Fourier transform of the time-dependent $\beta_2$ value is shown for both experiment and model. The shaded area around the experimental curve corresponds to the standard error from the three separate pump-probe scans. The phases of the three prominent frequency components are also plotted for experiment and model (diamond markers). The standard error of the extracted phases are smaller than the marker dimensions.}
 \end{figure}
 
The observed time-dependent PAD of the 4$f$ manifold can be compared to the model $l$-uncoupling simulation. Our model employs spectroscopic values for the 4$f$ electronic energies in the vibrational ground state of N\textsubscript{2} \cite{huber_f_1994}, rather than the approximate form given by Eq. (\ref{simplified_H}). The amplitude and phase of the discrete Fourier transform of the time-dependent $\beta _2$ values from both experiment and model are shown in Fig. \ref{Fig4}. The model reproduces the main quantum beat frequencies seen in the experiment, which occur in the 0.7-1.5 THz (700 fs-1.4 ps) range. The extracted phases of the prominent frequency components are also plotted. The modeled phase values agree quite well with the measured phases, suggesting that the impulsive 5-photon excitation model provides an adequate description of the pump process. The discrepancy in phases amount to shifts of $<170$ fs, close to the duration of the excitation pulse and the limits of the impulsive model. Simulations of the higher order $\beta$ parameters show lower frequency oscillations ($<0.5$ THz) that are not present in the experimental signal. This discrepancy may be due to coupling to nearby perturbing electronic states in N\textsubscript{2} \cite{huber_f_1994,jungen_2003}, angular distortion inherent to non-inverted VMI images, or from the non-perturbative nature of the pump pulse \cite{althorpe_molecular_1999}.

In summary, we have used time-resolved photoelectron angular distributions to image the dynamics of the non-Born-Oppenheimer $l$-uncoupling regime in the lab frame. The complex behavior of the electron motion in this regime is radically different from low-lying valence state molecular frame motion. Future studies in which the ion-core and photoelectron angular distributions are simultaneously measured would be highly beneficial in allowing for a direct experimental comparison of the rotational and electronic motion of the $l$-uncoupled system. In addition, the coherent excitation of $l$-uncoupled electronic states shown in this work, when combined with existing rotational wave packet preparation techniques \cite{korobenko_adiabatic_2016,Staplefeldt_Siedeman}, offers the prospect of creating unique and exotic electronic-rotational wave packets not possible in Born-Oppenheimer systems.

\begin{acknowledgments}
This work was supported by the AMOS program, Chemical Sciences, Geosciences, and Biosciences Division, Basic Energy Sciences, Office of Science, U.S. Department of Energy.
\end{acknowledgments}

\bibliography{l_uncoupling.bib}

\end{document}